\documentclass[a4paper,twocolumn,10pt
]{quantumarticle}
\pdfoutput=1
\usepackage[utf8]{inputenc}
\usepackage[english]{babel}
\usepackage[T1]{fontenc}
\usepackage{amsmath,amssymb}
\usepackage{hyperref}

\usepackage{physics,xfrac}
\usepackage[nameinlink]{cleveref}

\newcommand{\sgn}{\text{sgn}}

\begin{document}

\title{Is the photon-blockade breakdown a quantum effect?\newline A neoclassical story}

\author{Árpád Kurkó}
\affiliation{HUN-REN Wigner RCP, H-1525 Budapest, P.O. Box 49., Hungary}
\author{Nikolett Német}
\affiliation{HUN-REN Wigner RCP, H-1525 Budapest, P.O. Box 49., Hungary}
\author{András Vukics}
\email{vukics.andras@wigner.hu}
\affiliation{HUN-REN Wigner RCP, H-1525 Budapest, P.O. Box 49., Hungary}


\begin{abstract}
The photon-blockade breakdown bistability can be intuitively explained invoking the energy spectrum of the interacting qubit-mode system. Yet, the neoclassical solution of the driven–dissipative Jaynes–Cummings model has been shown to capture several key aspects of the phenomenon. In this paper, we set out to compare a fully quantum solution with the neo- and semiclassical solutions. Although the neoclassical theory is founded on the assumption of a pure partial state for the qubit, 
it is not simply the $\gamma\to0$ limit of the semiclassical theory; the semi- vs. neoclassical duality being a case of non-commutativity of limits. Furthermore, we show that the neoclassical predictions still hold in case of a small qubit decay. Tracing the bistable behavior for different detunings, we show that it is robust over a significant range of $\Delta$ values. We demonstrate that the aptitude of the neoclassical description is founded on the high quantum purity of the bright state of the photon-blockade breakdown bistability, which sharply differentiates this phenomenology from conventional optical bistability. It is thereby demonstrated that driven–dissipative dynamics can produce closely separable pure steady states in an interacting bipartite.
\end{abstract}

\maketitle

\section{Introduction}

Optical bistability (OB) \cite{abraham1982optical,lugiato1984ii,reinisch1994optical} is a striking manifestation of the nonlinear nature of light-matter interaction. It was initially observed in macroscopic nonlinear media, like saturable resonators \cite{szoke1969bistable} and semiconductors \cite{gibbs1979optical}. In principle, however, a few atoms or even only a single atom could exhibit the characteristics of optical bistability \cite{savage1988single,dombi2013optical}. The progress in cavity QED experiments towards strong coupling between atoms and the cavity field, enabled the demonstration of bistability in this regime \cite{rempe1991optical,elsasser2004optical,sauer2004cavity,geng2020universal}. The nonlinear input–output relation  underlying this effect can be adequately described by the Maxwell–Bloch equations, that are conveniently derived as the mean-field representation of the driven–dissipative Jaynes–Cummings model. This description will be referred to as the \emph{semiclassical} model for the purposes of this paper.

Much more recently, the breakdown of photon blockade, another phenomenology that involves intensity bistability in an “extremely strong” coupling regime of an interacting qubit-mode system described by the driven–dissipative Jaynes–Cummings model has been proposed \cite{alsing1991spontaneous,dombi2015bistability,carmichael2015breakdown,gutierrez2018dissipative} and experimentally observed on the circuit QED platform \cite{fink2017observation,fitzpatrick2017observation,sett2022emergent}. It has been shown that in a strong-coupling limit, in analogy with a thermodynamic limit \cite{carmichael2015breakdown,vukics2019finite,curtis2021critical}, the photon-blockade breakdown (PBB) turns into a first-order dissipative phase transition, which is a recent paradigm in quantum science and technology \cite{DiCandia2021,petrovnin2023microwave}.

An intuitive picture of the PBB phenomenology goes as follows: the energy levels of the interacting qubit-mode system suffer Rabi splitting \cite{rempe1987observation, fink2008climbing}, due to which a drive tuned close to resonance with the bare frequency of the subsystems cannot excite the system prepared in the ground state—the photon blockade \cite{imamoḡlu1997strongly, birnbaum2005photon, faraon2008coherent, lang2011observation}. However, for any such drive detuning, there exists a region in one of the Rabi subladders where the ladder spacing matches this detuning. Such a region can accommodate a closely coherent state, that can be reached by a combination of multi-photon transitions \cite{kubanek2008two, shamailov2010multi, dombi2015bistability} and photon-number increasing quantum jumps if the drive is strong enough—the breakdown of photon blockade.

On the theoretical front, in contrast to OB with its semiclassical backdrop, PBB was shown to be related to the \emph{neo}classical Jaynes–Cummings theory \cite{carmichael2015breakdown}. In the above intuitive explanation, the full quantum mechanical energy spectrum of the interacting bipartite system in a driven–dissipative setting plays a crucial role. Nevertheless, the boundaries of the domain of bistability (shape of the phase diagram) in the parameter space and the strong-coupling thermodynamic limit are correctly predicted by the neoclassical theory.

While the semiclassical model, in its various orders of the cumulant expansion, has been widely used in the study of light-matter interaction, neoclassical theory has remained somewhat of an undercurrent for decades. In contrast to the semiclassical theory, which is applicable in the case of vanishing qubit decay only in a restricted sense (more on this below), the neoclassical theory \emph{presupposes} $\gamma=0$. This involves the conservation of the length of the qubit pseudospin, which is \emph{equivalent} to the purity of the partial state of the qubit. This entails that the qubit cannot be entangled with the mode,\footnote{Strictly speaking, this inference applies only when the full bipartite state is pure, but, as we will demonstrate, this largely holds in the studied scenario.} which is at least surprising for an interacting quantum system. Naively, one would believe that the two theories coincide in the $\gamma\to0$ limit, but this is not the case: the $t\to\infty$ limit (steady state) and the $\gamma\to0$ limit (vanishing qubit decay) are not interchangeable here.

Although the neoclassical theory has found indirect justification in the PBB scenario in the shape of the phase diagram and the nature of the thermodynamic limit, its fundamental assumption, the conservation of the length of the pseudospin has not been directly tackled. It is highly nontrivial what a real interacting and driven-dissipative quantum system would do in the $\gamma=0$, or, even more interestingly, small $\gamma$ case.

In this paper we directly study the validity of this fundamental neoclassical assumption – no entanglement – along the temporal evolution of quantum trajectories over a wide variety of parameter sets. We clarify why this assumption remains valid to a good approximation even after the addition of a small qubit decay. The prediction of the parameter boundaries of bistability remains robust, reinforcing the neoclassical theory as the correct classical backdrop of the PBB phenomenology. In \cref{sec:JCclassical} we review the two kinds of classical models, pointing out the intricacies of the $\gamma\to0$ limit. Subsequently, in \cref{sec:Quantum} we compare their predictions to fully quantum trajectory solutions in the cases of vanishing and small but finite qubit decay. Finally, in \cref{sec:Effective}, we present a model for the instantaneous quantum states along the PBB bistability that explains the overall aptitude of the neoclassical model, and the extent of necessary quantum corrections.

\section{The driven-dissipative\newline Jaynes–Cummings model\newline and its classical descendants}
\label{sec:JCclassical}
Our discussion of the coupled system of a single qubit and a single mode, including decay of the two subsystems is based on the master equation (we set $\hbar = 1$)
\begin{subequations}
\label{eq:master}
\begin{multline}
\label{eq:Liouvillian}
    \Dot{\rho} = -i\,[H, \rho] + \kappa\,\qty( 2a\rho a^\dag - a^\dag a\,\rho - \rho\,a^\dag a )\\+ \gamma\,\qty( 2\sigma\rho \sigma^\dag - \sigma^\dag \sigma \rho - \rho\,\sigma^\dag \sigma ) \, ,
\end{multline}
where $a$ is the annihilation operator of the mode, $\sigma$ is the qubit lowering operator, $\kappa$ and $\gamma$ are the respective damping rates. The Hamiltonian is the coherently driven Jaynes–Cummings model:
\begin{equation}
\label{eq:RotFrameDrivenJC}
    H = -\Delta \qty(a^\dag a + \sigma^\dag \sigma ) + ig\, \qty( a^\dag \sigma - a\sigma^\dag ) + i\eta \, \qty( a^\dag - a ) \, ,
\end{equation}
\end{subequations}
where the mode and the qubit are resonant, $g$ is the coupling constant (single-photon Rabi frequency), and $\Delta,\;\eta$ are the drive detuning and amplitude, respectively.

The easiest way to obtain dynamical equations for expectation values is taking the time derivative of $\Tr\qty{\rho\,\mathcal O}$ to obtain $\Tr\qty{\dot\rho\,\mathcal O}$ and substitute $\dot\rho$ from \cref{eq:master}. After factorizing the expectation values of operator products, a complete set of equations is obtained for the mode operator, the qubit polarization, and the population:
\begin{subequations}
\label{eq:HeisEqs}
\begin{align}
    \dot{\expval{a}} =& \qty( i\Delta - \kappa )\expval{a} + g\expval{\sigma} + \eta \, ,\label{eq:HeisEqsA}\\
    \dot{\expval{\sigma}} =& \qty( i\Delta - \gamma )\expval{\sigma} + g\expval{a}\expval{\sigma_z}\, , \label{eq:HeisEqsB} \\
    \dot{\expval{\sigma_z}} =& -2g\,\text{Re}\qty{\expval{a^\dag}\expval{\sigma}} - \gamma(\expval{\sigma_z}+1) \, , \label{eq:HeisEqsC}
\end{align}
\end{subequations}
where the $\sigma_z$ Pauli operator is defined as $\sigma^\dag \sigma - \sigma \sigma^\dag$.

\subsection{The semiclassical theory}
The semiclassical solution is obtained by zeroing the LHS of \cref{eq:HeisEqs} (steady state, or $t\to\infty$ limit), whereupon the resulting system can be solved for an implicit equation for the photon number. In the resulting equation, the $\gamma\to0$ limit can be safely performed to obtain:
\begin{equation}
    \expval{n} = \frac{\eta^2}{\kappa^2 + \Delta^2\,\qty[ 1 - \frac{g^2}{\Delta^2+2g^2\expval{n}} ]^2} \, ,
    \label{eq:nSemi}
\end{equation}
cf. \cref{app:semi} for details of this calculation.

\begin{figure}
\centering
\includegraphics[width=\linewidth]{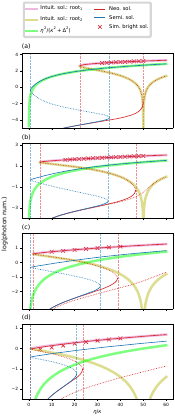}
\caption{The real roots (the dotted parts correspond to nonphysical solutions) of the self-consistent equations for the photon number in the neoclassical (\ref{eq:nNeo}) and semiclassical (\ref{eq:nSemi}) theories (red and blue curves), for different drive detunings: $\Delta=\lbrace 2, 10, 25, 50\rbrace$ (from top to bottom), and for $\gamma=0, g=100\kappa$. The red (blue) dashed vertical lines indicate the boundaries of the bistable regions for the neoclassical (semiclassical) solution. The red 'x' markers represent results from quantum-jump Monte Carlo simulations for the average photon number during the bright periods. The two real roots from the “intuitive” picture (cf. \cref{app:intuit}) are plotted as pink and yellow transparent lines; and the detuned empty-cavity photon number $\eta^2/\qty(\Delta^2+\kappa^2)$ as a green transparent line.}
\label{fig:phNumClMCS}
\end{figure}

This equation is equivalent to a cubic equation, so that the roots can be found analytically. The real roots are plotted in \cref{fig:phNumClMCS} as a function of the drive amplitude for four different values of the detuning. Throughout this article, we consider a strong coupling constant, $\kappa \ll g = 100 \kappa$, and a driving field with $\Delta > 0$. For a given detuning, two different domains can be distinguished based on the number of roots. There is a domain with three real roots (one of which is nonphysical since it involves decreasing photon numbers with increasing drive amplitudes), which is surrounded by single-root regions. It is apparent that the lower boundary is at very small $\eta$ values and is largely independent of $\Delta$, whereas the bistability region shrinks with increasing detuning, as the upper boundary is decreasing. The photon number of the “bright” state is close to that of the detuned empty mode, $\eta^2/\qty(\Delta^2+\kappa^2 )$, the difference being noticeable only for relatively large detunings as in \cref{fig:phNumClMCS}(c,d). Close to the resonance, the dispersive shift $g^2/\qty( \Delta^2+2g^2\expval{n})$ is almost negligible due to the very high bright-state photon numbers. As we increase $\Delta$, the detuned empty-mode photon number underestimates the semiclassical solution due to the increasing significance of the dispersion, which pushes the mode towards resonance.

We close the discussion of the semiclassical model by noting that the bistability in the $\gamma\to0$ limit is very fragile here, since it vanishes by the addition of even an infinitesimal amount of phase noise for the qubit, cf. \cref{app:semi}.

\subsection{The neoclassical theory}
What if we attempt the $\gamma\to0$ limit first? In that case, we immediately run into the problem that \cref{eq:HeisEqsC} becomes degenerate when applied to the steady state: it is just the homogeneous equation $\text{Re}\qty{\expval{a^\dag}\expval{\sigma}}=0$.

On the other hand, one can notice that with $\gamma=0$, the system \labelcref{eq:HeisEqs} conserves the length of the pseudospin defined as
\begin{subequations}
\begin{equation}
\mathcal{S} \equiv 4\abs{\expval{\sigma}}^2+\expval{\sigma_z}^2.
\end{equation}
It is also straightforward to verify that $\mathcal{S}$ is an affine function of the purity of the partial state of the qubit:
\begin{equation}
\mathcal{S} = 2\Tr\qty{\rho_\text{qubit}^2}-1,\qqtext{with}\rho_\text{qubit}=\Tr_\text{mode}\qty{\rho}.
\end{equation}
\end{subequations}
This implies that a separable initial state (e.g. the ground state) remains separable throughout the time evolution.

In the following, it will become apparent that in the very strong-coupling regime, in particular, for the PBB phenomenology, the neoclassical one is the correct classical theory. Even its fundamental assumption, the constraint on the pseudospin, that is, the purity of the qubit partial state—separability of the full state can hold to a good approximation in a fully quantum model, cf. \cref{sec:Quantum}. It is intriguing why a strongly interacting quantum dynamics fails to create entanglement between the parts. We attempt to answer this question in \cref{sec:Effective}.

From the neoclassical theory, the implicit equation for the expectation value of the photon number reads
\begin{equation}
    \expval{n} = \frac{\eta^2}{\kappa^2 + \qty[ \Delta - \frac{\sgn(\Delta)g^2}{\sqrt{4g^2\expval{n}+\Delta^2}}]^2} \, ,
\label{eq:nNeo}
\end{equation}
which now leads to a fourth-order equation in $\expval{n}$. For this reason, four real solutions are possible, of which the two extreme remain real over the full range of $\eta$. However, the lowest-lying solution is always non-physical in this case, as we demonstrate in \cref{app:intuit}.

Typical solutions are displayed in \cref{fig:phNumClMCS}. The superiority of the neoclassical theory over the semiclassical one is already apparent in the prediction of the bright-state photon number by the former: it closely aligns with the fully quantum trajectory results, with differences appearing only for large detunings $\Delta/g \gtrsim 1/4$.

\begin{figure}
\centering
\includegraphics[width=\linewidth]{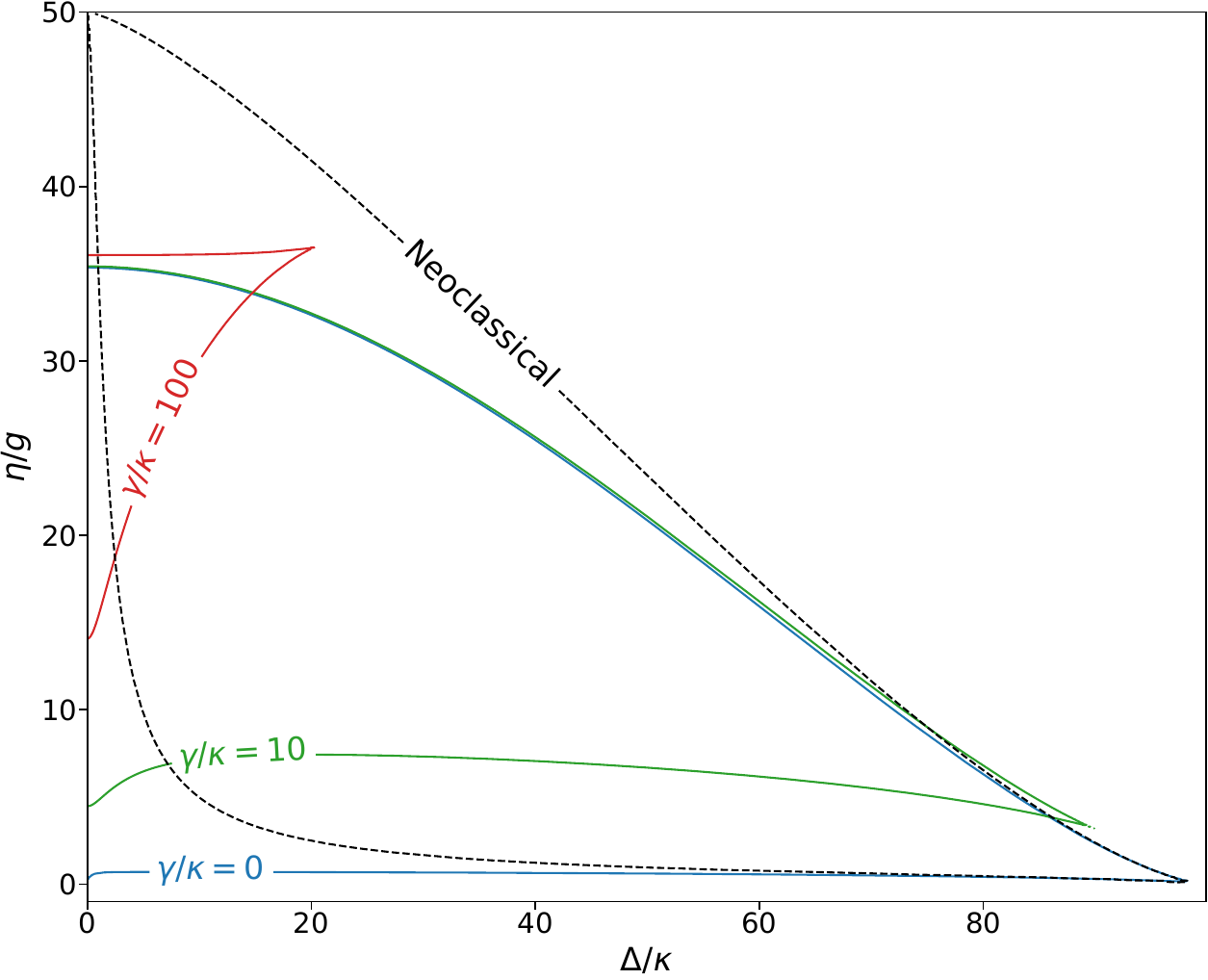}
\caption{Boundaries of the bistable region predicted by the semiclassical theory for different values of qubit decay rate $\gamma$ compared to the boundary predicted by the neoclassical model. The coupling constant is $g=100\kappa$.}
\label{fig:SemiNeoBound}
\end{figure}

In the same figure, the significant difference between the bistability regions predicted by the two classical theories is also apparent. This is further exposed in \cref{fig:SemiNeoBound}, where we moreover show the impact of increasing $\gamma$ on the semiclassical solution. The semiclassical “dim” region increases with increasing $\gamma$, while the upper boundaries of the bistable region remain nearly unchanged until very strong qubit decay.

Comparing the neoclassical bistability region with the $\gamma=0$ semiclassical one, the most striking difference is the finite bistability interval at $\Delta=0$ predicted by the the latter, the semiclassical upper boundary going to the empty-mode photon number: $\eta^2/\kappa^2$, cf. \cref{eq:nSemi}, whereas in the neoclassical theory, the bistability region shrinks into a point (a “critical point” in the phase-transition language, cf. \cite{alsing1991spontaneous,carmichael2015breakdown}) at $\eta=g/2$. It is only for large detunings that the two bistability regions overlap to a large extent: as $\Delta$ increases, the semiclassical boundary converges to the neoclassical one. This trend is supported by the $\Delta$-dependence of the pseudospin derived from the semiclassical equations. As the detuning varies $0\to\infty$, the semiclassically predicted pseudospin\footnote{For the sake of completeness, we quote the length of the pseudospin derived from the semiclassical theory: $$\mathcal{S} = 1 - \qty(\frac{2g^2\expval{n}}{2g^2\expval{n} + \Delta^2})^2$$.} changes $0\to1$, hence, for large values of $\Delta$ the two classical theories become equivalent.

\section{Quantum-trajectory simulations}
\label{sec:Quantum}
The quantum-jump Monte Carlo trajectories for the purposes of this section are generated with the C++QED framework \cite{vukics2007cpp,vukics2012cpp,sandner2014cpp}. We consider two values for the qubit decay rate: (i) $\gamma=0$, and (ii) $\gamma=0.01\kappa$, the second choice being made to move slightly away from the key condition of the neoclassical theory.

In earlier works exploring the PBB regime, quantum trajectories confirmed the bistable behavior in the time domain \cite{dombi2015bistability,carmichael2015breakdown,fink2017observation,vukics2019finite}. Telegraph signals emerge with alternating dim and bright  periods. We define the filling factor $\mathcal{F}$ as the ration of the total time spent by the system in the bright period and the full simulation time. In the language of the steady-state density operator, it is equivalent to the mixing ratio of the two states: $\rho_\text{ss} = (1-\mathcal F)\,\rho_\text{dim} + \mathcal F \,\rho_\text{bright}$. For a given detuning, $\mathcal F$ grows $0\to1$ with increasing drive strength between the boundaries of the bistability region.\footnote{Approaching the boundaries is difficult due to the long simulation times necessary for sufficient statistics on the very rare occurrences of the bright (dim) state close to the lower (upper) boundary. Therefore, in order to compare with the classical predictions of the shape of the bistability domain we work close to half-filling $\mathcal{F}=0.5$.}

\begin{figure}
\centering
\begin{tabular}{l}
(a) \\
\includegraphics[width=\linewidth]{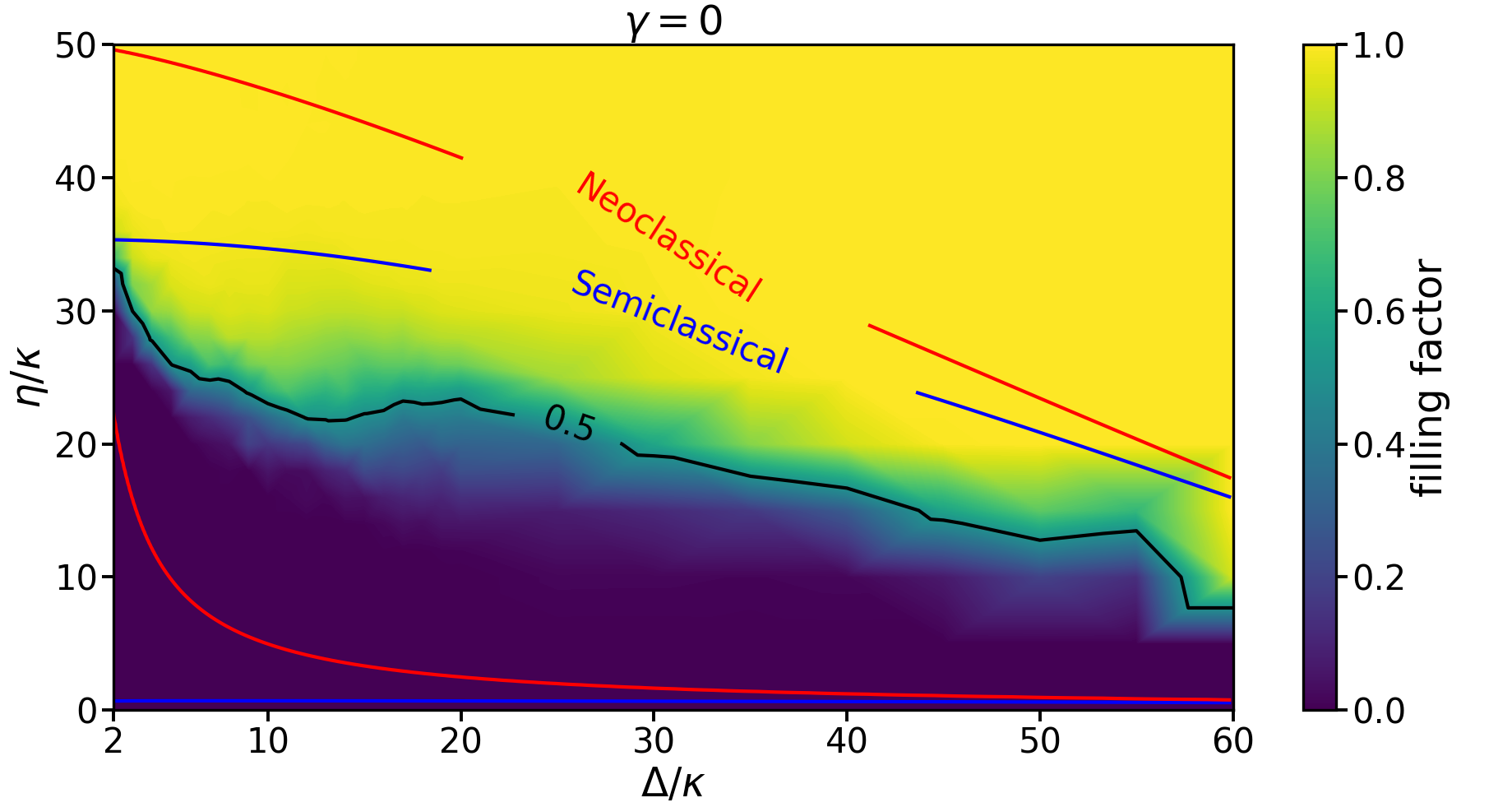} \\
(b) \\
\includegraphics[width=\linewidth]{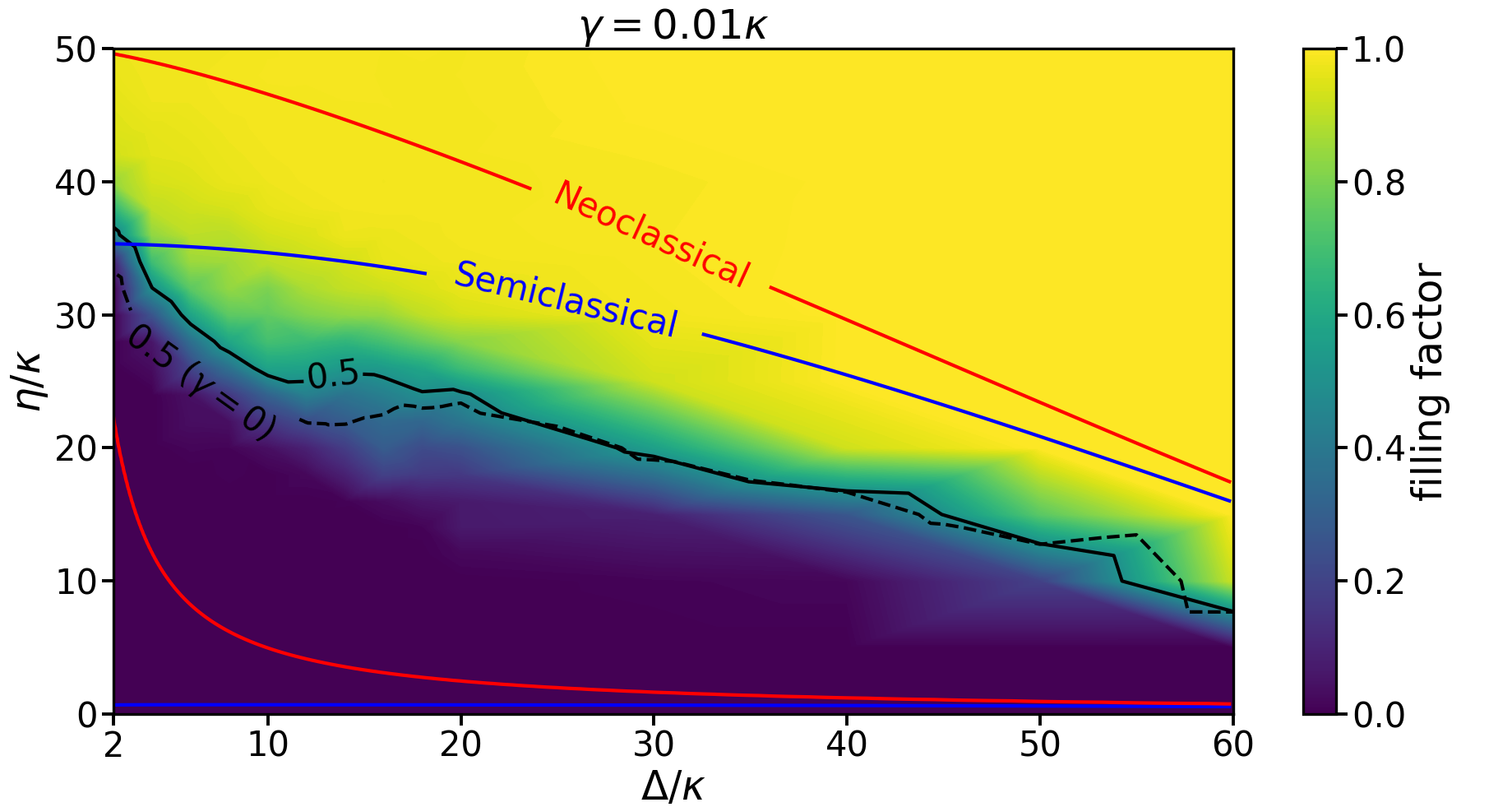}
\end{tabular}
\caption{Filling factor on the $\Delta-\eta$ plane obtained from quantum trajectory simulations for $g=100\kappa$ and (a) $\gamma = 0$ (b) $\gamma = 0.01\kappa$. The drive amplitude resulting in half-filling of the telegraph signals is plotted in black curve (in panel (b), this curve from panel (a) is repeated as a dashed line for comparison). The boundaries of the bistable region predicted by the classical theories are plotted for comparison, the red neoclassical curves being repeated on panel (b), as they do not depend on $\gamma$.}
\label{fig:Filling}
\end{figure}

To compare the quantum-trajectory results with the predictions of the classical theories, we plot $\mathcal F$ over the $\Delta-\eta$ space in \cref{fig:Filling}. While in earlier works, the driving was close to resonance, here, before anything else, we can draw the conclusion that the bistability in the time domain remains robust up until the relatively large detunings $\Delta\lesssim60\kappa$.\footnote{While the focus of the present paper is different, it’s worth mentioning that dwell times in the attractors are significantly affected by increasing either $\Delta$ or $\gamma$, resulting in decreases in both cases.}

For small detunings $\Delta \lesssim g/4$, where the difference between the boundaries predicted by the semiclassical and neoclassical theories is significant, the neoclassical solution aligns manifestly better with the quantum results. Choosing $\gamma=0.01\kappa$ to move slightly away from zero qubit decay, cf.~\cref{fig:Filling}(b), the quantum trajectory result remains consistent with the neoclassical solution, whereas the semiclassical solution fails spectacularly, as its upper boundary falls below even the half-filling curve for small detunings.

\begin{figure}
\centering
\includegraphics[width=\linewidth]{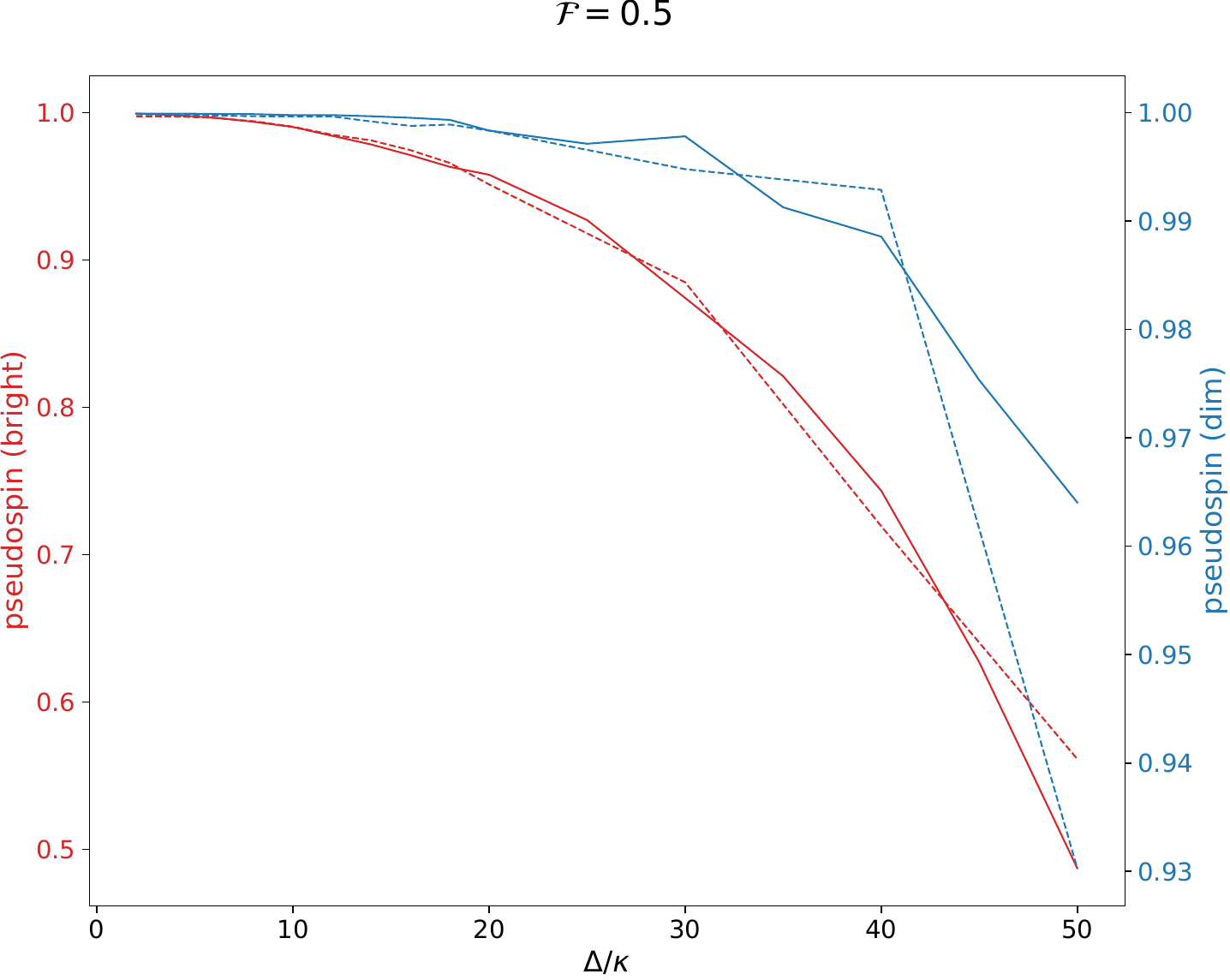}
\caption{The pseudospin in the bright (red) and dim (blue) periods of the telegraph signals with half-filling for $g=100\kappa$. The continuous curves (dashed) correspond to $\gamma=0$ ($\gamma=0.01\kappa$).}
\label{fig:PSpin}
\end{figure}

We now turn to a direct study of the basic assumption of the neoclassical theory, $\mathcal S=1$, along the quantum trajectories. \Cref{fig:PSpin} illustrates that this condition is verified to a good approximation for moderate detunings, independently of the presence of a small qubit decay. For increasing $\Delta$, the pseudospin of both the dim and the bright state decreases, but the latter more strongly, since the dim state remains close to $\ket{g,0}$ throughout, which is a pure separable state. The deviation of the bright state from $\mathcal{S} = 1$ on the quantum trajectories underlies the fact that the neoclassical theory overestimates the bright-state photon number for large detunings $\Delta = g/4$ and $g/2$, cf.~\cref{fig:phNumClMCS}(c,d).

Since the full bipartite qubit-mode state along a single trajectory is pure, the pseudospin of the qubit serves as an entanglement measure. Therefore, the results summarized in \cref{fig:PSpin} justify two of our claims made in the introduction regarding the aptitude of the neoclassical theory in the strong-coupling regime of the strongly-driven Jaynes–Cummings model:
\begin{enumerate}
\item The qubit–mode interaction does not generate entanglement over quantum trajectories in a significant parameter domain.
\item The neoclassical theory remains robust against the introduction of a small qubit decay.
\end{enumerate}

We proceed by introducing an effective model of the bright states that is able to explain the observed trends in entanglement.

\section{Effective model for the bright state}
\label{sec:Effective}
The dressed states of the undriven Jaynes–Cummings Hamiltonian, $\ket{m, \pm} = \frac{1}{\sqrt{2}}\qty( \ket{g,m} \pm \ket{e,m-1} )$ with $m = 1, 2, ...$, form two distinct, well-resolved anharmonic subladders in the strong-coupling regime. The strong anharmonicity is the basis of the photon blockade: for a drive not too far from resonance, $\Delta \ll g$, the system cannot be excited from the $\ket{g,0}$ ground state. On the other hand, for every such detuning value, there exists a region in the spectrum where one of the subladders (depending on the sign of $\Delta$) is closely resonant with the drive. This region is higher as the drive gets closer to resonance. In this region, a bright state can live, which can be attained by strong driving via a combination of multi-photon events and photon-number increasing quantum jumps – this is the breakdown of photon blockade.

With this picture in mind, we can attempt a model of the bright state, which is a pure state living in the ‘+’ Jaynes–Cummings subladder (since we have restricted ourselves to $\Delta>0$):
\begin{equation}
    \ket{\Psi_\text{bright}} = \sum_{m=0}^{\infty} c_m \ket{m,+}\, .
    \label{eq:AnsatzBright}
\end{equation}
For a concrete form below, we will assume that the coefficients $c_m$ depend only on the bright-state photon number $\expval{n}_\text{bright}$.

Let us examine the pseudospin length and entanglement degree of the bright state under the Ansatz \labelcref{eq:AnsatzBright}. It is easy to see that
\begin{subequations}
\begin{align}
\expval{\sigma_z} &= \Tr{ \sigma_z \rho} = -\abs{c_0}^2\qqtext{and}\\
\expval{\sigma} &= \Tr{\sigma \rho} = \frac12\sum_{m=0}^{\infty} c_{m+1}c_m^* \equiv \frac12\, \mathcal{C}.
\end{align}
\end{subequations}
The length of the pseudospin is then:
\begin{equation}
    \mathcal{S} = \expval{\sigma_z}^2 + 4\abs{\expval{\sigma}}^2 = \abs{c_0}^4 + \abs{\mathcal{C}}^2 \, .
    \label{eq:PSpinDressed}
\end{equation}

As the value of $\expval{n}_\text{bright}$ increases, the dimension of the region in the Jaynes–Cummings subladder occupied by the system during the bright periods also increases, because the harmonicity of the ladder increases with increasing height. The simplest way to quantify this process is to assume that the $c_m$ amplitudes approach a uniform distribution, making that $\mathcal C=\sum_{m=0}^{\infty} c_{m+1}c_m^*\approx\sum_{m=0}^{\infty} \abs{c_m}^2 = 1$, whereas $c_0\to0$.\footnote{It is remarkable that this model of the bright state predicts a completely saturated qubit.} This results in $\mathcal S=1$, the fulfillment of the neoclassical condition, reproducing the small-$\Delta$ (i.e. large photon number) behavior in \cref{fig:PSpin}.

\begin{figure*}
\centering
\includegraphics[width=.8\linewidth]{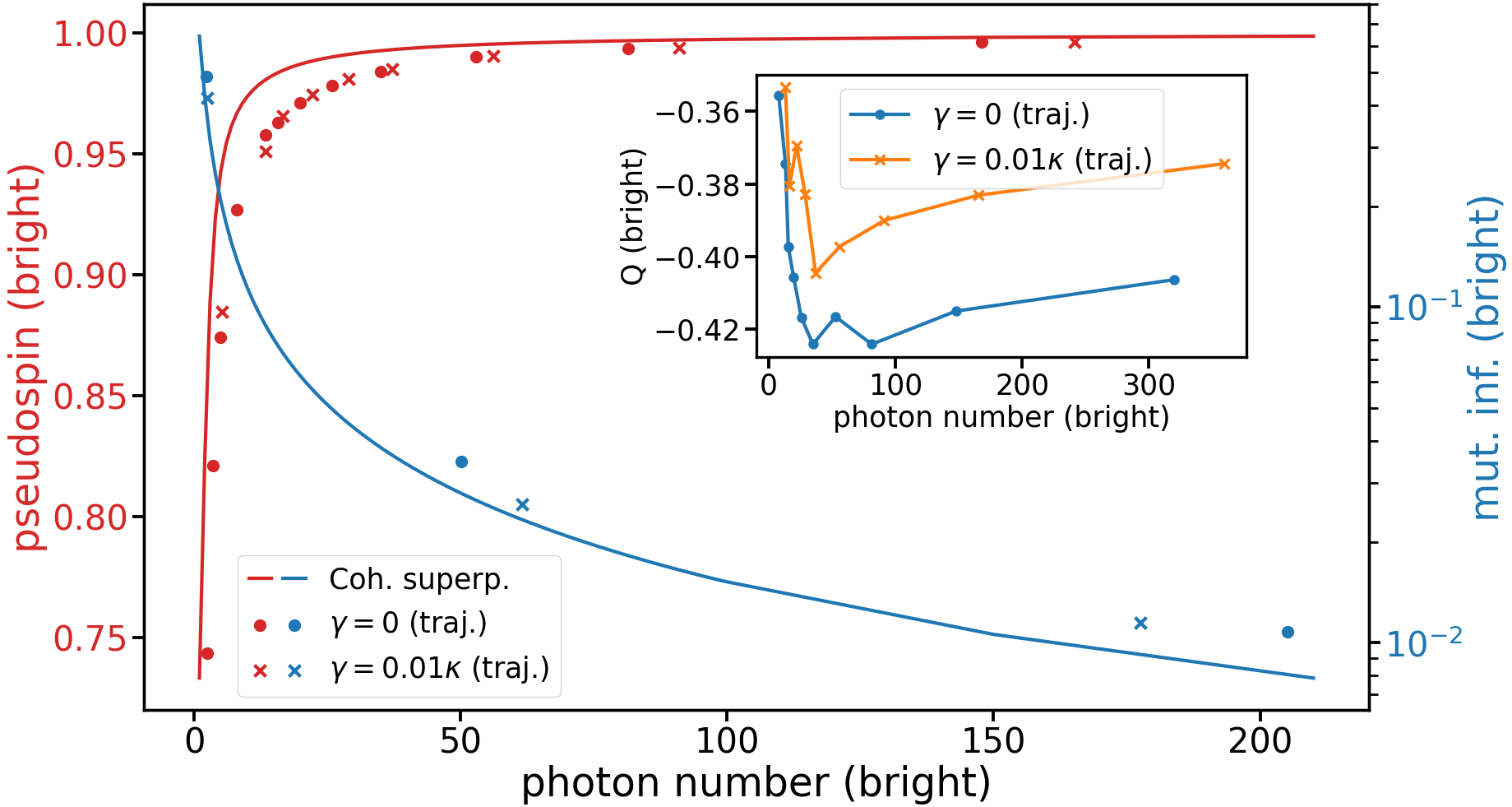}
\caption{The pseudospin of the qubit (red curve) and the quantum mutual information of the qubit-mode system (blue curve) in terms of the photon number of the bright state, considering the bright state as a coherent superposition \labelcref{eq:PoissonSuperp} of the Jaynes–Cummings dressed states. The entanglement measures derived from the trajectory simulations are also presented: the $\gamma=0$ plotted as circle markers,  $\gamma=0.01\kappa$ points plotted as 'x' markers. The inset displays the Mandel-Q parameter of the bright state in terms of the bright-state photon number from the trajectory simulations: $\gamma=0$ points plotted as blue circle markers, $\gamma=0.01\kappa$ points plotted as orange 'x' markers.}
\label{fig:CoherentSuperposition}
\end{figure*}

In the following, in order to concretize the implications of the Ansatz \labelcref{eq:AnsatzBright}, we take the $c_m$s to be the coefficients of a coherent state. This is a natural assumption given the competition of drive and dissipation on a closely equidistant segment of the ‘+’ subladder. The coefficients read
\begin{equation}
    c_m = \exp{ -\frac{\expval{n}_\text{bright}}2} \sqrt{\frac{\expval{n}_\text{bright}^m}{m!}} \, .
    \label{eq:PoissonSuperp}
\end{equation}
In \cref{fig:CoherentSuperposition} we display the dependence of the pseudospin (and the entanglement) on $\expval{n}_\text{bright}$ obtained by plugging the coefficients \labelcref{eq:PoissonSuperp} back into \cref{eq:PSpinDressed}. The length of the pseudospin remains close to 1 as the bright-state photon number decreases $\sim200\to20$, whereupon $\mathcal{S}$ decreases abruptly. The entanglement follows a complementary curve.\footnote{Turning around the “deviation of $\mathcal S$ from 1” notion, we obtain a more direct quantification of the entanglement in the bright state: the \emph{quantum mutual information}. It is defined as
$I_{\text Q : \text M} = S(\rho_\text{Q}) + S(\rho_\text{M}) - S(\rho)$, with the reduced density operator $\rho_\text{Q(M)}$ for the qubit (mode) subsystem and the von Neumann entropy $S(\rho)=-\Tr\qty{\rho\log\rho}$. As the full qubit-mode system is pure on a single trajectory, the mutual information simplifies to $I_{\text Q : \text M} = 2S(\rho_\text{Q})$. This measure has a minimum value of zero, which corresponds to a separable state, and a maximum value of $2\log\qty[ \text{dim}(\mathcal{H}_\text{Q}) ] = 2$, implying that $\ket{\Psi_\text{bright}}$ is maximally entangled. The Ansatz \labelcref{eq:AnsatzBright} results in the mutual information $I_{\text Q : \text M} = -2\qty( \lambda_1\log\lambda_1  + \lambda_2\log\lambda_2 )$, where the eigenvalues of $\rho_\text{Q}$ read
\begin{equation*}
     \lambda_{1,2} = \frac{2-\abs{c_0}^2 \pm \sqrt{\abs{c_0}^4+4\abs{\mathcal{C}}^2}}{4} \, .
    \label{eq:EigenvalDressed}
\end{equation*}
In the limit of large bright-state photon number, that is, $\mathcal C\to1$ and $c_0\to0$, the eigenvalues do $\lambda_1\to1$ and $\lambda_2\to0$, which makes the mutual information vanish, indicating a separable qubit-mode system.}

In the same figure, we display data points extracted from the full quantum trajectory simulations presented in \cref{sec:Quantum}. Even though in these simulations we have varied 3 parameters ($\gamma$, $\Delta$, and $\eta$), when plotted only as a function of $\expval{n}_\text{bright}$, the points fit very well on the model curves, reinforcing that the bright state is close to the form \labelcref{eq:AnsatzBright} with the coefficients \labelcref{eq:PoissonSuperp}.

Differences between the simulation data points and the model curves are maximal for photon numbers between 10 and 50. As the inset of \cref{fig:CoherentSuperposition} shows, this is also where the Mandel-Q parameter \footnote{The Mandel-Q parameter is defined as the normalized variance of the photon number operator: $Q\equiv\text{var}\qty(n)/\expval{n} - 1$} of the bright state is maximal, indicating that the system occupies the lower-lying, anharmonic part of the spectrum. Both the low- and high-photon-number bright states are close to being classical, but for different reasons: the former because they are close to the vacuum $\ket{0}$, which is itself a coherent state, the latter because of the state being close to the form \labelcref{eq:AnsatzBright} with the coefficients \labelcref{eq:PoissonSuperp}. In between these two extremes, the state deviates from a coherent state, and hence the deviation from the Ansatz \labelcref{eq:AnsatzBright} is maximal. We furthermore note that since the driven harmonic oscillator is the only quantum system whose state remains pure in the presence of dissipation, the full bipartite state is expected to become mixed when the bright state falls to the strongly anharmonic part of the spectrum. However, the purity of the full qubit-mode state remains a subject for further study.

\section{Conclusions}
\label{sec:conclusion}
This paper was motivated by the intriguing adequacy of the Jaynes–Cummings neoclassical theory at reproducing some aspects of the photon-blockade breakdown phenomenology, i.e. the boundaries of the bistability region and the thermodynamic limit. This is in spite of the very strong assumption of the neoclassical theory: the conservation of the pseudospin, which in the quantum case is equivalent to the qubit being in a pure state. Among other things this means that there can be no entanglement between the mode and the qubit, which is at least surprising for an interacting quantum system.

The two bistable phenomena, optical bistability, known since long ago, and the more recent photon-blockade breakdown have different classical backdrops: the semi- and the neoclassical theories, respectively. Naively, one could believe that the two will coincide in the $\gamma\to0$ limit, but this is not the case due to the constraint on the length of the pseudospin in the neoclassical theory. Quantum mechanically, this means that in the bright state, the state of the qubit is pure, whereas its state is mixed in conventional optical bistability. It is non-trivial what a real (i.e. fully quantum interacting driven-dissipative) system would do in the $\gamma=0$, or the small $\gamma$ case – this latter is even more intriguing, since the neoclassical condition can not be exactly fulfilled.

In this paper we have demonstrated in the language of quantum-trajectory simulations that the neoclassical theory is the correct classical theory of the photon-blockade breakdown phenomenology, which does not break down immediately by the addition of a small $\gamma$. Assuming that the bright state is a coherent state in one of the Jaynes–Cummings subladders, we could reproduce the behavior of the length of the pseudospin and the qubit-mode entanglement as a function of the photon number.

As an outlook, we finally note that our results touch upon the difference between conventional optical bistability and photon-blockade breakdown: whereas in the former, the bright state is necessarily mixed, in the latter it can be very close to a pure state. We have hence demonstrated that driven–dissipative dynamics can produce closely separable pure steady states in an interacting bipartite, that can have interesting consequences in dissipative quantum state preparation \cite{verstraete2009quantum,harrington2022engineered}.

\section{Acknowledgment}
This research was supported by the Ministry of Culture and Innovation and the National Research, Development and Innovation Office within the Quantum Information National Laboratory of Hungary (Grant No. 2022-2.1.1-NL-2022-00004), and the ERANET COFUND QuantERA program  (MOCA, 2019-2.1.7-ERA-NET-2022-00041).
We are grateful to the HUN-REN Cloud (\url{http://science-cloud.hu}) for providing us with the
suitable computational infrastructure for the simulations. We thank Peter Domokos for stimulating discussions and valuable remarks on the manuscript.

\appendix

\section{The semiclassical solution and its sensitivity to phase noise}
\label{app:semi}
Here we outline the derivation of the semiclassical solution, i.e. the steady-state photon number of the Maxwell–Bloch equations. We go beyond the minimum decay rate $\gamma$ of \cref{eq:HeisEqs}, which is needed for consistency with spontaneous emission. Hence, we include an additional coherence decay rate $\gamma_{\text c}$ by adding some extra dephasing term to the equation of the qubit polarization:
\begin{subequations}
\label{eq:HeisEqsDeph}
\begin{align}
    0 =& \qty(i\Delta-\kappa)\expval{a} + g\expval{\sigma} + \eta \, ,\label{eq:HeisEqsDephA}\\
    0 =& \qty(i\Delta-\gamma_{\perp})\expval{\sigma} + g\expval{a}\expval{\sigma_z}\, , \label{eq:HeisEqsDephB} \\
    0 =& -2g\,\text{Re}\qty{\expval{a^\dag}\expval{\sigma}} - \gamma(\expval{\sigma_z}+1) \, , \label{eq:HeisEqsDephC}
\end{align}
\end{subequations}
where the transverse damping rate is defined as $\gamma_{\perp} \equiv \gamma + \gamma_{\text c}$. The self-consistent equation for the cavity intensity using \cref{eq:HeisEqsDephA} reads
\begin{equation}
    \expval{n} = \frac{\eta^2}{\abs{i\Delta + \kappa + S(\expval{n})}^2} \, ,
    \label{eq:PhNum}
\end{equation}
where the dispersive shift $S(\expval{n})$ is $g\expval{\sigma}/\expval{a}$. After eliminating the qubit polarization and population using \cref{eq:HeisEqsDephB,eq:HeisEqsDephC}, we obtain
\begin{equation}
    S(\expval{n}) = -\frac{g^2\gamma\qty(i\Delta+\gamma_\perp)}{2g^2\expval{n}\gamma_\perp + \gamma\qty(\Delta^2+\gamma_\perp^2)} \, .
    \label{eq:DispShift}
\end{equation}
At this point we investigate the $\gamma \to 0$ limit for two scenarios, with and without phase noise. Having a vanishing $\gamma_\text{c}$, i.e. not considering the dephasing effect of atom-atom collisions, the transverse decay rate is purely consisted of the natural damping rate. As a result, $\gamma$ appears as a common factor in the nominator and denominator of \cref{eq:DispShift}. Thus, in the case of zero phase noise 
\begin{equation}
    S_{\gamma_\text{c} = 0}(\expval{n}) = -\frac{g^2 \qty(i\Delta+\gamma)}{2g^2\expval{n}+\Delta^2+\gamma^2} \, ,
    \label{eq:DispShift0Deph}
\end{equation}
we can safely take the $\gamma \to 0$ limit, and recover \cref{eq:nSemi} for the photon number. However, that is not the case when there is some phase noise in the system, as the dispersive shift vanishes for $\gamma_\perp \to \gamma_\text{c}$, cf. \cref{eq:DispShift}. This means that in the $\gamma \to 0$ limit the bistability in the semiclassical model is very fragile as we introduce a small amount of phase noise for the qubit. 

Last point to mention regarding the semiclassial solution, is the algebraic equivalence of the absorptive and dispersive bistability. This becomes apparent examining the form of the dispersive shift with $\gamma_\text{c} = 0$, cf. \cref{eq:DispShift0Deph}, where we observe complete reciprocity between $\Delta$ and $\gamma$.

\section{An intuitive estimation for the photon number}
\label{app:intuit}
The Jaynes–Cummings model is equivalent to two nonlinear oscillators, with the energy separation $\omega_\text{M} \pm \qty(\sqrt{n+1}-\sqrt{n})$\footnote{$\omega_\text{M}$ is the angular frequency of the mode} of the adjacent dressed states $\qty{\ket{n+1,\pm}, \ket{n, \pm} }$. In the large photon number limit $n \gg 1$, where the ladders become nearly equidistant, the separations can be approximated as $\omega_\text{M} \pm g/\qty(2\sqrt{n})$. Where the drive detuning closely matches the spacing of one of the ladders, a coherent state with the photon number
\begin{equation}
    \label{eq:phNumInt}
    n' = \frac{\eta^2}{\kappa^2 + \qty(\Delta - \frac{g}{2\sqrt{n'}})^2}
\end{equation}
is created. Considering a positive detuning, we have taken into account the positive excitation path in \cref{eq:phNumInt}, i.e. the drive is off-resonant with the other ladder $\ket{n,-}$. As \cref{eq:phNumInt} is quadratic in $\sqrt{n'}$, a well-defined value of $\eta$ exists for each $\Delta$, separating the region where no real solutions exist from the region with two distinct real roots. These critical values of the drive amplitude coincide with the lower boundary of the bistability region predicted by the neoclassical theory, as shown in \cref{fig:phNumClMCS}. In the domain of the two real solutions, the first root returns the neoclassical bright state photon number, this can be observed in \cref{fig:phNumClMCS} and \cref{eq:nNeo}, which in the $4g^2\expval{n} \gg \Delta^2$ limit can be approximated by \cref{eq:phNumInt}. The second real root of \cref{eq:phNumInt} zeros at $\eta=g/2$ and gives a negative solution above it. This non-physical solution, i.e. $\sqrt{n'} < 0$, for small detunings overlaps in a wide range with the lowest-lying root of the neoclassical theory, cf. \cref{fig:phNumClMCS}.

\bibliographystyle{quantum}
\bibliography{PBB}

\begin{thebibliography}{10}

\bibitem{abraham1982optical}
E~Abraham and SD~Smith.
\newblock ``Optical bistability and related devices''.
\newblock \href{https://dx.doi.org/10.1088/0034-4885/45/8/001}{Reports on
  Progress in Physics {\bf 45}, 815}~(1982).

\bibitem{lugiato1984ii}
Luigi~A Lugiato.
\newblock ``Theory of optical bistability''.
\newblock In Progress in optics.
\newblock \href{https://dx.doi.org/10.1016/S0079-6638(08)70122-7}{Volume~21,
  chapter~II, pages 69--216}.
\newblock Elsevier~(1984).

\bibitem{reinisch1994optical}
R~Reinisch and G~Vitrant.
\newblock ``Optical bistability''.
\newblock \href{https://dx.doi.org/10.1016/0079-6727(94)90004-3}{Progress in
  quantum electronics {\bf 18}, 1--38}~(1994).

\bibitem{szoke1969bistable}
A~Sz{\"o}ke, V~Daneu, J~Goldhar, and NA~Kurnit.
\newblock ``Bistable optical element and its applications''.
\newblock \href{https://dx.doi.org/10.1063/1.1652866}{Applied Physics Letters
  {\bf 15}, 376--379}~(1969).

\bibitem{gibbs1979optical}
HM~Gibbs, SL~McCall, TNC Venkatesan, AC~Gossard, A~Passner, and W~Wiegmann.
\newblock ``Optical bistability in semiconductors''.
\newblock \href{https://dx.doi.org/10.1063/1.91157}{Applied Physics Letters
  {\bf 35}, 451--453}~(1979).

\bibitem{savage1988single}
CM~Savage and HJ~Carmichael.
\newblock ``Single atom optical bistability''.
\newblock \href{https://dx.doi.org/10.1109/3.7075}{IEEE journal of quantum
  electronics {\bf 24}, 1495--1498}~(1988).

\bibitem{dombi2013optical}
Andr{\'a}s Dombi, Andr{\'a}s Vukics, and Peter Domokos.
\newblock ``Optical bistability in strong-coupling cavity {QED} with a few
  atoms''.
\newblock \href{https://dx.doi.org/10.1088/0953-4075/46/22/224010}{Journal of
  Physics B: Atomic, Molecular and Optical Physics {\bf 46}, 224010}~(2013).

\bibitem{rempe1991optical}
G~Rempe, RJ~Thompson, RJ~Brecha, WD~Lee, and HJ~Kimble.
\newblock ``Optical bistability and photon statistics in cavity quantum
  electrodynamics''.
\newblock \href{https://dx.doi.org/10.1103/PhysRevLett.67.1727}{Physical Review
  Letters {\bf 67}, 1727}~(1991).

\bibitem{elsasser2004optical}
Th~Els{\"a}sser, B~Nagorny, and A~Hemmerich.
\newblock ``Optical bistability and collective behavior of atoms trapped in a
  high-q ring cavity''.
\newblock \href{https://dx.doi.org/10.1103/PhysRevA.69.033403}{Physical Review
  A {\bf 69}, 033403}~(2004).

\bibitem{sauer2004cavity}
JA~Sauer, KM~Fortier, MS~Chang, CD~Hamley, and MS~Chapman.
\newblock ``Cavity qed with optically transported atoms''.
\newblock \href{https://dx.doi.org/10.1103/PhysRevA.69.051804}{Physical Review
  A {\bf 69}, 051804}~(2004).

\bibitem{geng2020universal}
Z~Geng, KJH Peters, AAP Trichet, K~Malmir, R~Kolkowski, JM~Smith, and SRK
  Rodriguez.
\newblock ``Universal scaling in the dynamic hysteresis, and non-markovian
  dynamics, of a tunable optical cavity''.
\newblock \href{https://dx.doi.org/10.1103/PhysRevLett.124.153603}{Physical
  Review Letters {\bf 124}, 153603}~(2020).

\bibitem{alsing1991spontaneous}
P~Alsing and HJ~Carmichael.
\newblock ``Spontaneous dressed-state polarization of a coupled atom and cavity
  mode''.
\newblock \href{https://dx.doi.org/10.1088/0954-8998/3/1/003}{Quantum Optics:
  Journal of the European Optical Society Part B {\bf 3}, 13}~(1991).

\bibitem{dombi2015bistability}
Andr{\'a}s Dombi, Andr{\'a}s Vukics, and Peter Domokos.
\newblock ``Bistability effect in the extreme strong coupling regime of the
  jaynes-cummings model''.
\newblock \href{https://dx.doi.org/10.1140/epjd/e2015-50861-9}{The European
  Physical Journal D {\bf 69}, 1--8}~(2015).

\bibitem{carmichael2015breakdown}
HJ~Carmichael.
\newblock ``Breakdown of photon blockade: A dissipative quantum phase
  transition in zero dimensions''.
\newblock \href{https://dx.doi.org/10.1103/PhysRevX.5.031028}{Physical Review X
  {\bf 5}, 031028}~(2015).

\bibitem{gutierrez2018dissipative}
Ricardo Guti{\'e}rrez-J{\'a}uregui and HJ~Carmichael.
\newblock ``Dissipative quantum phase transitions of light in a generalized
  jaynes-cummings-rabi model''.
\newblock \href{https://dx.doi.org/10.1103/PhysRevA.98.023804}{Physical Review
  A {\bf 98}, 023804}~(2018).

\bibitem{fink2017observation}
Johannes~M Fink, Andr{\'a}s Dombi, Andr{\'a}s Vukics, Andreas Wallraff, and
  Peter Domokos.
\newblock ``Observation of the photon-blockade breakdown phase transition''.
\newblock \href{https://dx.doi.org/10.1103/PhysRevX.7.011012}{Physical Review X
  {\bf 7}, 011012}~(2017).

\bibitem{fitzpatrick2017observation}
Mattias Fitzpatrick, Neereja~M Sundaresan, Andy~CY Li, Jens Koch, and Andrew~A
  Houck.
\newblock ``Observation of a dissipative phase transition in a one-dimensional
  circuit qed lattice''.
\newblock \href{https://dx.doi.org/10.1103/PhysRevX.7.011016}{Physical Review X
  {\bf 7}, 011016}~(2017).

\bibitem{sett2022emergent}
Riya Sett, Farid Hassani, Duc Phan, Shabir Barzanjeh, Andras Vukics, and
  Johannes~M Fink.
\newblock ``Emergent macroscopic bistability induced by a single
  superconducting qubit''~(2022).

\bibitem{vukics2019finite}
Andr{\'a}s Vukics, Andr{\'a}s Dombi, Johannes~M Fink, and P{\'e}ter Domokos.
\newblock ``Finite-size scaling of the photon-blockade breakdown dissipative
  quantum phase transition''.
\newblock \href{https://dx.doi.org/10.22331/q-2019-06-03-150}{Quantum {\bf 3},
  150}~(2019).

\bibitem{curtis2021critical}
Jonathan~B. Curtis, Igor Boettcher, Jeremy~T. Young, Mohammad~F. Maghrebi,
  Howard Carmichael, Alexey~V. Gorshkov, and Michael Foss-Feig.
\newblock ``Critical theory for the breakdown of photon blockade''.
\newblock \href{https://dx.doi.org/10.1103/PhysRevResearch.3.023062}{Phys. Rev.
  Res. {\bf 3}, 023062}~(2021).

\bibitem{DiCandia2021}
R~Di~Candia, F~Minganti, KV~Petrovnin, GS~Paraoanu, and S~Felicetti.
\newblock ``Critical parametric quantum sensing''.
\newblock
  \href{https://dx.doi.org/https://doi.org/10.1038/s41534-023-00690-z}{npj
  Quantum Information {\bf 9}, 23}~(2023).

\bibitem{petrovnin2023microwave}
Kirill Petrovnin, Jiaming Wang, Michael Perelshtein, Pertti Hakonen, and
  Gheorghe~Sorin Paraoanu.
\newblock ``Microwave photon detection at parametric criticality''~(2023).

\bibitem{rempe1987observation}
Gerhard Rempe, Herbert Walther, and Norbert Klein.
\newblock ``Observation of quantum collapse and revival in a one-atom maser''.
\newblock \href{https://dx.doi.org/10.1103/PhysRevLett.58.353}{Physical review
  letters {\bf 58}, 353}~(1987).

\bibitem{fink2008climbing}
JM~Fink, M~G{\"o}ppl, M~Baur, R~Bianchetti, Peter~J Leek, Alexandre Blais, and
  Andreas Wallraff.
\newblock ``Climbing the jaynes--cummings ladder and observing its nonlinearity
  in a cavity qed system''.
\newblock \href{https://dx.doi.org/10.1038/nature07112}{Nature {\bf 454},
  315--318}~(2008).

\bibitem{imamoḡlu1997strongly}
A~Imamoḡlu, Helmut Schmidt, Gareth Woods, and Moshe Deutsch.
\newblock ``Strongly interacting photons in a nonlinear cavity''.
\newblock \href{https://dx.doi.org/10.1103/PhysRevLett.79.1467}{Physical Review
  Letters {\bf 79}, 1467}~(1997).

\bibitem{birnbaum2005photon}
Kevin~M Birnbaum, Andreea Boca, Russell Miller, Allen~D Boozer, Tracy~E
  Northup, and H~Jeff Kimble.
\newblock ``Photon blockade in an optical cavity with one trapped atom''.
\newblock \href{https://dx.doi.org/10.1038/nature03804}{Nature {\bf 436},
  87--90}~(2005).

\bibitem{faraon2008coherent}
Andrei Faraon, Ilya Fushman, Dirk Englund, Nick Stoltz, Pierre Petroff, and
  Jelena Vu{\v{c}}kovi{\'c}.
\newblock ``Coherent generation of non-classical light on a chip via
  photon-induced tunnelling and blockade''.
\newblock \href{https://dx.doi.org/10.1038/nphys1078}{Nature Physics {\bf 4},
  859--863}~(2008).

\bibitem{lang2011observation}
C~Lang, Deniz Bozyigit, Christopher Eichler, L~Steffen, JM~Fink,
  AA~Abdumalikov~Jr, M~Baur, Stefan Filipp, Marcus~P Da~Silva, Alexandre Blais,
  et~al.
\newblock ``Observation of resonant photon blockade at microwave frequencies
  using correlation function measurements''.
\newblock \href{https://dx.doi.org/10.1103/PhysRevLett.106.243601}{Physical
  review letters {\bf 106}, 243601}~(2011).

\bibitem{kubanek2008two}
Alexander Kubanek, Alexei Ourjoumtsev, Ingrid Schuster, Markus Koch, Pepijn~WH
  Pinkse, Karim Murr, and Gerhard Rempe.
\newblock ``Two-photon gateway in one-atom cavity quantum electrodynamics''.
\newblock \href{https://dx.doi.org/10.1103/PhysRevLett.101.203602}{Physical
  Review Letters {\bf 101}, 203602}~(2008).

\bibitem{shamailov2010multi}
SS~Shamailov, AS~Parkins, MJ~Collett, and HJ~Carmichael.
\newblock ``Multi-photon blockade and dressing of the dressed states''.
\newblock \href{https://dx.doi.org/10.1016/j.optcom.2009.10.062}{Optics
  Communications {\bf 283}, 766--772}~(2010).

\bibitem{vukics2007cpp}
A~Vukics and H~Ritsch.
\newblock ``{C++QED: an object-oriented framework for wave-function simulations
  of cavity QED systems}''.
\newblock
  \href{https://dx.doi.org/https://doi.org/10.1140/epjd/e2007-00210-x}{The
  European Physical Journal D {\bf 44}, 585--599}~(2007).

\bibitem{vukics2012cpp}
Andr{\'a}s Vukics.
\newblock ``{C++QEDv2: The multi-array concept and compile-time algorithms in
  the definition of composite quantum systems}''.
\newblock
  \href{https://dx.doi.org/https://doi.org/10.1016/j.cpc.2012.02.004}{Computer
  Physics Communications {\bf 183}, 1381--1396}~(2012).

\bibitem{sandner2014cpp}
Raimar Sandner and Andr{\'a}s Vukics.
\newblock ``{C++QEDv2 Milestone 10: A C++/Python application-programming
  framework for simulating open quantum dynamics}''.
\newblock
  \href{https://dx.doi.org/https://doi.org/10.1016/j.cpc.2014.04.011}{Computer
  Physics Communications {\bf 185}, 2380--2382}~(2014).

\bibitem{verstraete2009quantum}
Frank Verstraete, Michael~M Wolf, and J~Ignacio~Cirac.
\newblock ``Quantum computation and quantum-state engineering driven by
  dissipation''.
\newblock \href{https://dx.doi.org/https://doi.org/10.1038/nphys1342}{Nature
  physics {\bf 5}, 633--636}~(2009).

\bibitem{harrington2022engineered}
Patrick~M Harrington, Erich~J Mueller, and Kater~W Murch.
\newblock ``Engineered dissipation for quantum information science''.
\newblock
  \href{https://dx.doi.org/https://doi.org/10.1038/s42254-022-00494-8}{Nature
  Reviews Physics {\bf 4}, 660--671}~(2022).

\end{thebibliography}

\end{document}